\documentclass[prd,aps,secnumarabic,superscriptaddress,floatfix,preprintnumbers]{revtex4-1}
\usepackage[dvips]{graphicx,color}
\usepackage{amsmath}
\def\hh{\mbox{\rule{0pt}{12pt}}}

\def\Dot{\!\cdot\!}
\def\al{\alpha}
\def\be{\beta}
\def\ga{\gamma}
\def\de{\delta}
\def\De{\Delta}

\def\la{\lambda}

\def\ro{\rho}

\def\part{\partial}

\def\Ei{{\rm Ei}}

\newcommand{\mathsym}[1]{{}}
\newcommand{\unicode}[1]{{}}

\begin{document}

\title{Vacuum polarization corrections to the $n=2$ and $n=3$ levels in muonic $^4$He}

\author{Wayne W. Repko}\email{repko@pa.msu.edu}
\affiliation{Department of Physics and Astronomy, Michigan State University, East Lansing, Michigan 48824, USA}
\author{Stanley F. Radford}\email{sradford@brockport.edu}
\affiliation{Department of Physics, The College at Brockport, State
University of New York, Brockport, NY 14420}
\author{Duane A. Dicus}\email{dicus@physics.utexas.edu}
\affiliation{Department of Physics and Center for Particles and Fields,
University of Texas, Austin, TX 78712, USA}

\date{\today}

\begin{abstract}
The $n=2$ and $n=3$ levels for muonic Helium are calculated using a potential that includes all one-loop and recoil effects. Electronic vacuum polarization corrections are calculated using an extension of Kinoshita and Nio method. For $n=2$, the results are $2p_{1/2}-2s_{1/2}=1375.05\pm 1.4$ meV and $2p_{3/2}-2s_{1/2}=1521.65\pm 1.4$ meV, essentially in agreement with the latest summary of the current calculations. The $n=3$ results are summarized in tabular form and give $3p_{1/2}-3s_{1/2}=394.76\pm 0.43$ meV and $3d_{3/2}-3p_{3/2}=111.40$ meV.
\end{abstract}

\maketitle

\section{Introduction}

The tension between the determination of the proton size measured using muonic hydrogen energy levels \cite{nature,science} and electron scattering data \cite{Mohr} has focused renewed attention on accurate calculations of the level splitting in muonic atoms \cite{Borie:1982ax,Borie_1,Pachucki,Pachucki_1,Pohl,Peset}. This discrepancy is also being addressed by using increasingly precise measurements of the electronic energy  levels of ordinary hydrogen \cite{Beyer,Fleurbaey}. Part of the problem in the muonic hydrogen extraction of the proton radius is related to the appropriate modeling of its complicated electromagnetic structure consisting of both electric and magnetic form factors. There are also many more QED corrections since the presence of an additional spin means that there are hyperfine and tensor interactions as well as mixing between energy levels. While these corrections are well understood, they, in some sense, get in the way of figuring out the proper way to model the size parameters of the proton. As a consequence, studying a system without these additional terms could prove helpful in understanding the nuclear corrections. 

 The case of muonic $^4$He is considerably simpler  since the nucleus is spinless and there is only one nuclear form factor and a single spin \cite{Borie:2012zz,Diepold}. The underlying potential that describes a muon interacting with a point nucleus is
\begin{eqnarray}\label{V}
V(\vec{r})&=&-Z\al\left\{\frac{1}{r}-\frac{\pi}{2\,m_1^2}\de(\vec{r})- \left(\frac{1}{2\,m_1^2}+\frac{1}{m_1m_2}\right)\frac{\vec{L}\Dot\vec{S}}{r^3} +\frac{1}{2m_1m_2}p_i\left(\frac{\de_{ij}}{r}+\frac{\hat{x}_i\hat{x}_j}{r} \right)p_j\right. \nonumber \\
&&\left.\hspace{30pt}-\frac{\al}{2\,\pi}\left(\frac{1}{m_1^2}+\frac{1}{m_1\,m_2}\right) \frac{\vec{L}\Dot\vec{S}}{r^3}-\frac{4\al}{3}\ln(\mu/\la)\left(\frac{1}{m_1^2}+\frac{2Z}{m_1\,m_2} +\frac{Z^2}{m_2^2}\right)\de(\vec{r})\right. \nonumber \\ [6pt]
&&\left.\hspace{30pt}+\frac{4\al}{3}\left(\frac{1}{m_1^2}\ln(\eta_2)+\frac{Z^2}{m_2^2} \ln(\eta_1)\right) \de(\vec{r})-\frac{2Z\al}{m_1m_2} \left(\frac{m_1^2\ln(\eta_1)-m_2^2\ln(\eta_2)}{m_1^2-m_2^2}\right)\de(\vec{r})\right. \nonumber \\ [6pt] 
&&\left.\hspace{30pt}-\frac{7Z\al}{6\pi\,m_1\,m_2}\nabla^2\left(\frac{\ln(\mu\,r+\ga)}{r}\right) \right\}\,.
\end{eqnarray}
The first line in Eq.\,(\ref{V}) is the fine structure contribution to the potential derived from the one photon exchange diagram and is usually called the Breit-Pauli potential. Including the relativistic corrections to the kinetic energy completes the fine structure contribution. The remaining terms are one-loop corrections obtained from the muonium potential of Ref.\cite{GRS} by neglecting the terms containing the muon spin ($\vec{\sigma}_2/2$). Ideally, the spin-independent terms should be verified by calculating the one-loop correction to the Breit-Pauli potential. This potential will be the starting point for the calculation of corrections to both the $n=2$ and $n=3$ levels. The parameters are: $m_1=m_\mu=105.658372$ MeV, $m_2=m_{He}=3728.400128$ MeV and $\al=7.297352566\;10^{-3}$. $\mu$ is the reduced mass, $\la$ is the infra-red cutoff, $\eta_j=m_j/(m_1+m_2)$ and $\ga$ is the Euler constant. 

Explicit calculations are given for the $n=2$ case to establish that this approach reproduces the usual results for the $n=2$ Lamb shift. The results for $n=3$ follow in the same way and are summarized in tabular form. In either case, the largest corrections to Eq.\,(\ref{V}) are QED electron vacuum polarization corrections to the Coulomb interaction and modifications due to the nuclear form factor. These will be treated in the next two Sections. The next largest corrections are those from the fine structure portion of the potential and are calculated in Section \ref{FS}. The one-loop corrections are treated in Section \ref{one-loop} and the $n=3$ results are summarized in Section \ref{n=3 lev}. There are also  numerous smaller vacuum polarization corrections to the fine structure terms that are summarized in an Appendix.

\section{Vacuum polarization corrections}

Perhaps the simplest way to treat the vacuum polarization corrections to the Coulomb potential is to use the momentum space approach of Kinoshita and Nio \cite{KN}. They show that the order by order energy corrections can be expressed as
\begin{eqnarray} \label{DelE}
\De E_{n,\ell} &=& -\frac{4\pi\,Z\al}{(2\pi)^3}\int\!\!d^{\,3}k \,\frac{1}{\vec{k}^{\,2}}\left(\Pi^{(2)}_f(\vec{k}^{\,2}) +\Pi^{(2)}_f(\vec{k}^{\,2})\Pi^{(2)}_f(\vec{k}^{\,2})+\Pi^{(4)}_f(\vec{k}^{\,2}) \right.\nonumber \\ 
&&\left. +\Pi^{(2)}_f(\vec{k}^{\,2})\Pi^{(2)}_f(\vec{k}^{\,2})\Pi^{(2)}_f(\vec{k}^{\,2})+ 2\Pi^{(2)}_f(\vec{k}^{\,2})\Pi^{(4)}_f(\vec{k}^{\,2})+\Pi^{(6)}_f(\vec{k}^{\,2})+ \ldots \right)\ro_{n\ell}(k^2a^2)\,,
\end{eqnarray}
where $a$ denotes the Bohr radius $(\mu\al)^{-1}$ and $\ro_{n\ell}(k^2a^2)$ is the Fourier transform of the non-relativistic probability density
\begin{equation} \label{rho}
\ro_{n\ell}(k^2a^2)=\int\!\!d^{\,3}r\langle |\psi_{n\ell}^m(\vec{r})|^2\rangle e^{-i\vec{k}\cdot\vec{r}}\,.
\end{equation}
The angular brackets indicate an average over the degenerate $m$ values for a given $\ell$. For the $n=2$ states \cite{KN}
\begin{equation}\label{KNrho}
\ro_{2S}(k^2a^2)=\frac{Z^4(Z^4-3k^2a^2 Z^2+2k^4a^4)}
{(Z^2+k^2a^2)^4}\,, \qquad \ro_{2P}(k^2a^2) =\frac{Z^6(Z^2-k^2a^2)}{(Z^2+k^2a^2)^4}\,.
\end{equation}
Note that Eq.\,(\ref{DelE}) contains both reducible e.g., $\Pi^{(2)}_f(\vec{k}^{\,2})\Pi^{(2)}_f(\vec{k}^{\,2})$, and irreducible e.g., $\Pi^{(4)}_f(\vec{k}^{\,2})$, contributions. The sum of these contributions can be obtained if one uses the dispersion representation of K\"allen \cite{GK} and Lehmann \cite{HL}. The electron vacuum polarization corrections to the Coulomb potential take the form
\begin{equation} \label{corr}
D(\vec{k}^{\,2})=-\frac{Ze^2}{\vec{k}^{\,2}}-Ze^2\int_{4m_e^2}^\infty\!\!\frac{d\la}{\la} \frac{\De (\la)}{\la + \vec{k}^{\,2}}\,,
\end{equation}
where
\begin{equation}
\Delta(q^2)=-\frac{(2\pi)^3}{3q^2}\sum_n\de^{(4)}(q-q_n)\langle0|j_\mu(0)|n\rangle \langle n|j^\mu(0)|0\rangle\,.
\end{equation}
For any order in $e^2$, the integral in Eq.\,(\ref{corr}) results in 
\begin{equation} \label{Pin}
-Ze^2\int_{4m_e^2}^\infty\!\!\frac{d\la}{\la} \frac{\De^{(n)}(\la)}{\la + \vec{k}^{\,2}}=-\frac{Ze^2}{\vec{k}^{\,2}}\Pi^{(n)}(\vec{k}^{\,2})\,,
\end{equation}
where $\Pi^{(n)}(\vec{k}^{\,2})$ is the sum of all $n/2$ loop vacuum polarization corrections, both reducible and irreducible. In general, the expression for the vacuum polarization correction at order $n$ is
\begin{equation}
\De E_{VP}^{(n)}(n,\ell,Z)=-\frac{4\pi\,Z\al}{(2\pi)^3}\int_{4m_e^2}^\infty \!\!\frac{d\la}{\la} \De^{(n)}(\la)\int\! \frac{d^3k}{\vec{k}^2+\la} \rho_{n\,\ell}(k^2a^2)\,,
\end{equation}
For the $2s$ and $2p$ states, the $d^3k$ integrals are
\begin{equation}
\frac{\pi^2\,Z^3(2a^2\la+Z^2)}{2a(a\sqrt{\la}+Z)^4}\quad {\rm and}\quad \frac{\pi^2\,Z^5}{2a(a\sqrt{\la}+Z)^4}\,.
\end{equation}
With these results, the $2s$ and $2p$ states can be expressed as
\begin{eqnarray}
\De E_{VP}^{(n)}2S &=&  -\frac{\mu\,Z^4\al^2}{4} \int_4^\infty\! \frac{dx}{x}\De^{(n)}(x) \frac{(2\be^2x+Z^2)}{(\be\sqrt{x}+Z)^4}\,, \\
\De E_{VP}^{(n)}2P &=&  -\frac{\mu\,Z^6\al^2}{4} \int_4^\infty\! \frac{dx}{x}\De^{(n)}(x) \frac{1}{(\be\sqrt{x}+Z)^4}\,,
\end{eqnarray}
where $\la=m_e^2\,x$ and $\be=m_e\,a$.
In particular, for the $e\bar{e}$ intermediate state, $\De^{(2)}(x)$ is
\begin{equation}
\De^{(2)}(x)=\frac{\al}{3\pi}(1+2/x)\sqrt{1-4/x}\,\theta(x-4)\,.
\end{equation}
For the fourth order Kallen-Sabry \cite{KS} contribution, $\De^{(2)}(x)$ is replaced by $\De^{(4)}(\sqrt{x}/2)$, where $\De^{(4)}(x)$ is given in Ref.\cite{repko-dicus}. The results are contained in Table \ref{HeTot}. All numerical integrals were evaluated using the NIntegrate routine in Mathematica.

There is also a fourth order contribution to the electron vacuum polarization arising from second order perturbation theory. It can be written as
\begin{equation}
E^{(2)}_{n\ell}=\mu^2Z\al\int_0^\infty\!\!dr\,r^2R_{n\ell}(Zr/a)VP(r)\int_0^\infty\!\! dr'\,r'^2\,g_{n\ell}(Zr/a,Zr'/a)VP(r')R_{n\ell}(Zr'/a)\,,
\end{equation}
where $VP(r)$ is
\begin{equation}
VP(r)=-\frac{Z\al}{r}\int_4^\infty\!\!\frac{dy}{y}\Delta^{(2)}(y)e^{-m_e\sqrt{y}\,r}
\,,
\end{equation}
and $g_{n\ell}(Zr/a,Zr'/a)$ is the radial Coulomb Green's function. For $n=2$ and $\ell=0$ with $x=Zr/a,\,x'=Zr'/a$, $E^{(2)}_{20}$ is
\begin{equation}
E^{(2)}_{20}=\frac{\mu Z^2\al^2}{2}\!\!\int_4^\infty\!\! \frac{dy}{y}\Delta^{(2)}(y)\,\,\int_4^\infty\!\!\frac{dy'}{y'}\Delta^{(2)}(y')\, \,\int_0^\infty\!\!dx\,x(1-x/2)e^{-x/2}e^{-bx}\,\,\int_0^\infty\!\!
dx'\,x'g_{20}(x,x')e^{-b'x'}(1-x'/2)e^{-x'/2}\,.
\end{equation}
Here $b=m_ea\sqrt{y}/Z$ and $b'=m_ea\sqrt{y'}/Z$. The integrals over $x$ and $x'$ can be evaluated analytically and the remaining integrals were evaluated numerically. There is a similar expression for $E^{(2)}_{21}$ with $xe^{-x/2}$ replacing $(1-x/2)e^{-x/2}$ and $24$ replacing $2$. The radial Green's functions $g_{20}(x,x')$ and $g_{21}(x,x')$ are \cite{HP,Hostler,JH,LF,repko-dicus}
\begin{subequations}
\begin{eqnarray}
g_{20}(x,x') &=& Ze^{-(x+x')/2}\left[(2-x)(2-x')\left(\ln(x)+\ln(x')+\frac{(x+x')}{4}+2\ga-\frac{15}{4}-{\rm Ei}(x_<)\right)\right. \nonumber \\
&& \left.+12-2x-2x'-\frac{2}{x}-\frac{2}{x'}+\frac{x}{x'}+\frac{x'}{x}-xx' +(2-x_>)e^{x_<}\left(\frac{1}{x_<}-1\right)\right]\,, \\ [6pt]
g_{21}(x,x') &=& \frac{Z}{3}xx'e^{-(x+x')/2}\left[\ln(x)+\ln(x')+ \frac{(x+x')}{4} +2\ga-\frac{49}{12}-\frac{3}{x}-\frac{3}{x^2}-\frac{2}{x^3} \right.\nonumber \\ 
&&\left.-\frac{3}{x'}-\frac{3}{x'^{\,2}}-\frac{2}{x'^{\,3}}-{\rm Ei}(x_<) +e^{x_<}\left(\frac{1}{x_<}+\frac{1}{x_<^2} +\frac{2}{x_<^3}\right)\right]\,, 
\end{eqnarray}
\end{subequations}
where $x=Zr/a$ and $x'=Zr'/a$. $g_{20}(x,x')$ differs from Ref.\,\cite{LF}. The numerical results are in Table \ref{HeTot}.

Finally, there are relativistic corrections to the one-loop electron vacuum polarization contributions. These were calculated using the results of Ref. \cite{rel}.

\section{Nuclear size corrections}
For a spinless nucleus, the simplest nuclear size connection comes from expanding the charge form factor as
\begin{equation}
F(\vec{k}^2)\sim 1-\frac{1}{6}\langle r^2\rangle\vec{k}^2+\cdots\,,
\end{equation}
where $\langle r^2\rangle\equiv r_N^2$ is the nuclear mean squared charge radius. This leads to an effective potential
\begin{equation}
\De V_N(\vec{r})=\frac{2}{3}\pi Z\al r_N^2\de(\vec{r})\,,
\end{equation}
and an energy correction in the $2s$ state of
\begin{equation}
\De E_N2S=\frac{\mu^3Z^4\al^4 r_N^2}{12}= 105.33\,r_N^2\,{\rm meV\,fm^{-2}}\,,
\end{equation}
for $Z=2$. This result is known to over estimate the correction because any reasonable form factor will contain a term order $r_N^3/a^3$ with the opposite sign. To illustrate this, consider the form factor $F(\vec{k}^2)=(r_N^2\vec{k}^2/12+1)^{-2}$, which is often used to parameterize form factors. The corresponding form of $\De E_N2S$ is
\begin{equation}
\De E_N2S=\frac{\mu^3Z^4\al^4\,r_N^2}{12} \left(1-\frac{5Zr_N}{2\sqrt{3}a}\right)=105.33\,r_N^2(1-0.01097\,r_N)\,.
\end{equation}
The result for helium with $r_N=1.681\pm 0.004$ fm \cite{Diepold} is given in Table \ref{HeTot}. There is very little dependence on the power of the form factor, but the correction is of order $r_N/a$ and its coefficient could be introduced as a parameter. The majority of the error in the nuclear size correction comes from the uncertainty in $r_N$. The value for the nuclear polarizability is taken from Ref. \cite{Sick}.
\begin{center}
\begin{table}[h!]
\begin{tabular}{|l|r|r|r|r|r|}
\hline
  &$2s_{1/2}$ &$2p_{1/2}$ &$2p_{3/2}$ &$2p_{1/2}-2s_{1/2}$ &$2p_{3/2}-2s_{1/2}$  \\
\hline
\hh one loop $m_e$ Vac. Pol.& $-2077.25$ & $-411.46$ & $-411.46$ & $1665.79$
& $1665.79$ \\
\hh one-loop $m_e$ rel. VP  & $-0.85$ & $-0.32$ & $-0.04$ & $0.53$ & $0.81$ \\
\hh Kallen-Sabry Vac. Pol.  & $-15.24$& $-3.67$ & $-3.67$ & $11.57$ & $11.57$ \\
\hh 2nd order $m_e$ Vac. Pol. & $-1.90$ & $-0.19$ & $-0.19$ & $1.71$ & $1.71$ \\
\hh one loop $m_\mu$ Vac. Pol. & $-0.34$ & $0.00$ & $0.00$ & $0.34$ & $0.34$ \\
\hh one loop $m_\pi$ Vac. Pol. & $-0.26$ & $0.00$ & $0.00$ & $0.26$ & $0.26$ \\
\hh &  &  &   &  &  \\
\hh Nuclear size & $292.20\pm 1.4$ & $0.00$ & $0.00$ & $-292.20\pm 1.4$ & $-292.20\pm 1.4$ \\
\hh Nuclear size VP corr. & $2.32$ & $-0.02$ & $-0.02$ & $-2.34$ & $-2.34$ \\
\hh Nuclear polarizability & $-2.35\pm 0.13$ &  &  & $2.35\pm 0.13$ & $2.35\pm 0.13$ \\
\hh &  &  &  &  & \\
\hh Fine structure & $-183.30$ & $-183.00$ & $-37.43$ & $0.29$ & $145.86$ \\
\hh Fine structure VP corr. & $0.86$ & $-0.11$ & $0.02$ & $-0.97$ & $-0.84$ \\
\hh Fine structure VP 2nd order & $0.262$ & $-0.17$ & $-0.02$ & $-0.44$ & $-0.29$ \\
\hh &  &  &  &  &  \\
\hh Order $\al^5$ Spin-Orbit & $0.00$ & $-0.31$ & $0.16$ & $-0.31$ & $0.16$ \\
\hh Order $\al^5$ Lamb shift & $11.57$ & $0.03$ & $0.03$ & $-11.54$ & $-11.54$ \\
\hh & & & & & \\
\hh Total & $-1974.28\pm 1.4$ & $-599.23$ & $-452.64$ & $1375.05\pm 1.4$ & $1521.65\pm 1.4$ \\
\hline
\end{tabular}
\caption{Contributions to the $2p_j-2s_{1/2}$ splitting for muonic helium. The entries are in meV. \label{HeTot}}
\end{table}
\end{center}

\section{Fine structure corrections \label{FS}}

The contributions from the various fine structure terms are
\vspace{6pt}
\begin{eqnarray}
\frac{Z\al\pi}{2\,m_1^2}\langle 2s_{1/2}|\de(\vec{r})|2s_{1/2}\rangle 
&=&\frac{\mu^3Z^4\al^4}{16\,m_1^2} \label{fsdelta} \\ [6pt]
Z\al\left(\frac{1}{2\,m_1^2}+\frac{1}{m_1m_2}\right)\langle 2p_j| \frac{\vec{L}\Dot\vec{S}}{r^3}|2p_j\rangle &=& \frac{\mu^3Z^4\al^4}{48} \left(\frac{1}{2\,m_1^2}+\frac{1}{m_1m_2}\right)\left(j(j+1)-11/4\right)\label{fsls} \\ [6pt]
-\frac{Z\al}{2m_1m_2}\langle 2s|p_i\left(\frac{\de_{ij}}{r} +\frac{\hat{x}_i\hat{x}_j}{r} \right)p_j |2s\rangle &=& -\frac{3\mu^3Z^4\al^4}{16m_1m_2} \\ [6pt]
-\frac{Z\al}{2m_1m_2}\langle 2p|p_i\left(\frac{\de_{ij}}{r} +\frac{\hat{x}_i\hat{x}_j}{r} \right)p_j |2p\rangle &=& -\frac{\mu^3Z^4\al^4}{16m_1m_2} \\ [6pt]
-\frac{1}{8}\left(\frac{1}{m_1^3}+\frac{1}{m_2^3}\right)\langle 2s| (\vec{p}^{\,2})^2|2s\rangle &=& -\frac{13}{128}\mu^4Z^4\al^4 \left(\frac{1}{m_1^3}+\frac{1}{m_2^3}\right) \\ [6pt]
-\frac{1}{8}\left(\frac{1}{m_1^3}+\frac{1}{m_2^3}\right)\langle 2p| (\vec{p}^{\,2})^2|2p\rangle &=& -\frac{7}{384}\mu^4Z^4\al^4 \left(\frac{1}{m_1^3}+\frac{1}{m_2^3}\right)
\end{eqnarray}
\newpage
Using these results, the total fine structure contributions are
\begin{eqnarray}
E_{FS}(2s_{1/2}) &=& -\frac{13}{128}\mu^4Z^4\al^4 \left(\frac{1}{m_1^3}+\frac{1}{m_2^3}\right)-\frac{3\mu^3Z^4\al^4}{16m_1m_2} +\frac{\mu^3Z^4\al^4}{16\,m_1^2} \\ [4pt]
E_{FS}(2p_{1/2}) &=&  -\frac{7}{384}\mu^4Z^4\al^4 \left(\frac{1}{m_1^3}+\frac{1}{m_2^3}\right)-\frac{\mu^3Z^4\al^4}{24} \left(\frac{1}{2\,m_1^2}+\frac{1}{m_1m_2}\right)-\frac{\mu^3Z^4\al^4}{16m_1m_2} \\ [4pt]
E_{FS}(2p_{3/2}) &=& -\frac{7}{384}\mu^4Z^4\al^4 \left(\frac{1}{m_1^3}+\frac{1}{m_2^3}\right)+\frac{\mu^3Z^4\al^4}{48} \left(\frac{1}{2\,m_1^2}+\frac{1}{m_1m_2}\right)-\frac{\mu^3Z^4\al^4}{16m_1m_2}
\end{eqnarray}
The numerical results are given in Table \ref{HeTot}. Note that the $2p_{1/2}$ and $2s_{1/2}$ are not degenerate due to recoil effects.

\section{One-loop contributions \label{one-loop}}
The one-loop corrections consist of a spin-orbit contribution and a spin-independent contribution, both of order $Z^4\al^5$. The spin-orbit correction is
\begin{equation}
\langle 2p_j|Z\al a_\mu\left(\frac{1}{m_1^2}+\frac{1}{m_1m_2}\right) \frac{\vec{L}\Dot\vec{S}}{r^3}|2p_j\rangle = \frac{\mu^3Z^4\al^4a_\mu}{48} \left(\frac{1}{m_1^2}+\frac{1}{m_1m_2}\right)(j(j+1)-11/4)\,,
\end{equation}
where $a_\mu$ is the muon anomalous magnetic moment.

The matrix element of the remaining one-loop spin independent term $V'_{SI}$ is \cite{GRS}
\begin{eqnarray}
\langle n\ell|V'_{SI}|n\ell\rangle &=& \frac{\mu^3Z^4\al^5}{n^3\pi} \left\{\frac{4}{3}\left(\left(\log\left[\frac{1}{Z^2\al^2k_0[n,0]}\right]+\frac{5}{6}\right) \left(\frac{1}{m_1^2}+\frac{2Z}{m_1m_2}+\frac{Z^2}{m_2^2}\right)\right.\right. \nonumber \\
&&\left.\left. -\left(\frac{1}{m_1^2}\log(\eta_2)+\frac{Z^2}{m_2^2}\log(\eta_1)\right)\right)+\frac{14Z}{3m_1m_2} \Big(\log\left[\frac{2Z\al}{n}\right]+\frac{n-1}{2n}+\sum_{j=0}^n\frac{1}{j}\Big)\right. \nonumber \\
&&\left.+\frac{2Z}{m_1m_2}\left(\frac{m_1^2\log(\eta_1)-m_2^2\log(\eta_2)}{(m_1^2-m_2^2)}\right) \right\}\de_{\ell 0} \nonumber \\
&&+\frac{\mu^3Z^4\al^5}{n^3\pi} \left\{\frac{4}{3}\log\left[\frac{1}{k_0[n,\ell]}\right]\left(\frac{1}{m_1^2}+\frac{2Z}{m_1m_2} +\frac{Z^2}{m_2^2}\right)-\frac{7Z}{3m_1m_2}\frac{1}{\ell(\ell+1)(2\ell+1)}\right\} (1-\de_{\ell 0})
\end{eqnarray}

The Bethe sums $k_0[n,\ell]$ used are $\ln(k_0[1,0])=2.9841285$, $\ln(k_0[2,0])=2.8117699$ and $\ln(k_0[2,1])=-0.0300167$ \cite{KM}.
Again, the numerical results are listed in Table \ref{HeTot}.

\section{Results for $n=3$ \label{n=3 lev}}

The corresponding results for $n=3$ can be calculated using the same techniques that were used in the $n=2$ case. Wave functions for the relativistic corrections can be derived from Rose \cite{Rose}, Eqs.\,(5.49a) and (5.49b). The necessary radial Green's functions are \cite{LF}
\begin{subequations}
\begin{eqnarray}
g_{30}(x,x')&=&\frac{4Z}{9}e^{-(x+x')/2}\left[\left(x^2/2-3x+3\right)\left(x'^2/2-3x'+3\right) \left(\log(x)+\log(x')+(x+x')/6+2\ga-25/6-\Ei(x_<)\right)\right.\nonumber \\
&& \left.+\left((3-x)x/3-5x/2-1/x+7\right)\left(x'^2/2-3x'+3\right)+\left(x^2/2-3x+3\right) \left((3-x')x'/3-5x'/2-1/x'+7\right)\right. \nonumber \\ [4pt]
&& \left. +\left(x_>^2/2-3x_>+3\right) e^{x_<}\left(x_</2+1/x_<-5/2\right)\right]\,, \\
g_{31}(x,x')&=&\frac{Z}{18}xx'e^{-(x+x')/2}\left[(4-x)(4-x') \left(\log(x)+\log(x')+(x+x')/6+2\ga-55/12-\Ei(x_<)\right)\right.\nonumber \\
&&\left.(4-x)\left(x'/3-6/x'-4/x'^2-2/x'^3+5\right)+ \left(x/3-6/x-4/x^2-2/x^3+5\right)(4-x')\right. \nonumber \\ [4pt]
&&\left.+(4-x_>)e^{x_<}\left(2/x_<^3+2/x_<^2+3/x_<-1\right)\right]\,, \\
g_{32}(x,x')&=&\frac{Z}{90}x^2x'^2e^{-(x+x')/2}\left[\log(x)+\log(x')+(x+x')/6+2\ga-89/20-\Ei(x_<) \right.\nonumber \\
&& \left.-5/x-10/x^2-20/x^3-30/x^4-24/x^5-5/x'-10/x'^2-20/x'^3-30/x'^4-24/x'^5 \right.\nonumber \\
&&\left. +e^{x_<}\left(1/x_<+1/x_<^2+2/x_<^3+6/x_<^4+24/x_<^5\right)\right]\,,
\end{eqnarray}
\end{subequations}
where $x=2Zr/3a$ and $x'=2Zr'/3a$. The $n=3$ Bethe sums can be found in Ref.\cite{KM}

The resulting level corrections and splittings are summarized in Table \ref{n=3}.
\begin{table}[h]
\centering
\begin{tabular}{|l|r|r|r|r|r|r|r|}
\hline
  &$3s_{1/2}$ &$3p_{1/2}$ &$3p_{3/2}$ &$3d_{3/2}$ &$3d_{5/2}$ &$3p_{1/2}-3s_{1/2}$ &$3d_{3/2}-3p_{3/2}$  \\
\hline
\hh one loop $m_e$ Vac. Pol.& $-603.07$ & $-121.83$ & $-121.83$ &$-11.37$ &$-11.37$ &$481.24$ &$110.46$ \\
\hh one-loop $m_e$ rel. VP  & $-0.26$ &$-0.10$ &$-0.02$ &$0.00$ &$0.00$ & $0.16$ &$0.02$ \\
\hh Kallen-Sabry Vac. Pol.  & $-4.41$ &$-1.05$ & $-1.05$ &$-0.13$ &$-0.13$ & $3.36$ &$0.92$ \\
\hh 2nd order $m_e$ Vac. Pol. & $-0.86$ &$-0.10$ &$-0.10$ &$0.00$ &$0.00$ &$0.76$ &$0.10$ \\
\hh one loop $m_\mu$ Vac. Pol. &$-0.10$ &$0.00$ &$0.00$ &$0.00$ &$0.00$ & $0.10$ & $0.00$ \\
\hh one loop $m_\pi$ Vac. Pol. &$-0.08$ &$0.00$ &$0.00$ &$0.00$ &$0.00$ & $0.08$ & $0.00$ \\
\hh &  &  &   &  &  &  &  \\
\hh Nuclear size &$86.58\pm 0.41$ &$0.00$ &$0.00$ &$0.00$ &$0.00$ & $-86.58\pm 0.41$ & $0.00$ \\
\hh Nuclear size VP corr. & $1.04$ &$-0.01$ &$-0.01$ &$0.00$ &$0.00$ & $-1.05$ & $0.01$ \\
\hh Nuclear polarizability &$-0.70\pm 0.13$  & 0.00 & 0.00 & 0.00 &0.00  & $0.70\pm 0.13$ & $ $ \\
\hh &  &  &  &  &  &  &  \\
\hh Fine structure & $-65.01$ &$-64.92$ & $-21.79$ &$-21.84$ &$-7.39$ & $0.09$ & $-0.05$ \\
\hh Fine structure VP corr. & $0.25$ &$-0.06$ &$-0.01$ &$0.00$ &$0.00$ & $-0.31$ &$0.01$ \\
\hh Fine structure VP 2nd order & $0.20$ &$-0.09$ &$-0.01$ &$0.00$ &$0.00$ & $-0.29$ & $0.01$ \\
\hh &  &  &  &  &  &  & \\
\hh Order $\al^5$ Spin-Orbit & $0.00$ & $-0.07$ &$0.03$ &$-0.02$ &$0.01$ & $-0.07$ & $-0.05$ \\
\hh Order $\al^5$ Lamb shift & $3.45$ &$0.01$ &$0.01$ &$0.00$ &$0.00$ & $-3.44$ &$-0.01$ \\
\hh & & & & & &  &\\
\hh Total & $-582.97\pm 0.43$ &$-188.20$ &$-144.76$ &$-33.37$ &$-18.88$ &$394.76\pm 0.43$ & $111.40$ \\
\hline
\end{tabular}
\caption{Contributions to the $3p_{1/2}-3s_{1/2}$ and $3d_{3/2}-3p_{3/2}$ splittings for muonic helium. The entries are in meV. \label{n=3}}
\end{table}
The nuclear polarizability correction was obtained by rescaling the result in Ref.\cite{Sick} using the $3s$ wave function evaluated at the origin.

\section{Conclusions}

Starting from a potential describing the QED interaction between a point scalar nucleus and a muon that includes the one-loop level corrections and all recoil effects, the Lamb shifts for $n=2$ and $n=3$ muonic helium have been calculated. The effects of electron vacuum polarization were included by extending the Kinoshita-Nio \cite{KN} method to all operators in the potential. This significantly simplifies the calculation of the various corrections. The nuclear size correction was modeled using a simple one-parameter form factor. Given an accurate value of $r_N$, the resulting correction can be included in the prediction of the $2p_{1/2}-2s_{1/2}$ and $2p_{3/2}-2s_{1/2}$ splittings and the corresponding results for $n=3$, or the correction can be used to extract a value of $r_N$ from spectroscopic data.

\appendix 
\section{Details of the Vacuum Polarization Corrections to the Fine Structure}

The additional corrections to the fine structure potential involve vacuum polarization modifications to the $1/r$ factors contained in first line of Eq.\,(\ref{V}) as well as traditional second order perturbative corrections. For the former, the approach of Ref.\cite{KN} is very convenient. Using this technique, the expectation value of any potential $V(r)$ can be expressed as
\begin{equation}\label{KNexp}
\langle n\ell |V|n\ell\rangle=\frac{1}{(2\pi)^3} \int\!d^3k\,V(k^2)\rho_{n\ell}(k^2a^2)\,,
\end{equation}
where $V(k^2)$ is the Fourier transform of $V(r)$ and $a=(\mu\al)^{-1}$. 

To see how this works, consider the fine structure contribution from the term $Z\al\pi\de(\vec{r})/(2m_1^2)$. Since the Fourier transform of $\de(\vec{r})$ is 1, the contribution to the energy is
\begin{equation}\label{del0}
\langle n\ell|\,\frac{Z\al\pi}{2m_1^2}\de(\vec{r})\,|n\ell\rangle = \frac{\mu^3\,Z\al^4}{4\pi m_1^2}\int_0^\infty\!\!dy\,y^2\rho_{n\ell}(y)\,,
\end{equation}
where $y=ka$. Using Eqs.\,(\ref{KNrho}), this gives Eq.\,(\ref{fsdelta}) for the $2s$ state and $0$ for the $2p$ state. To include the effect of electron vacuum polarization, recall that the leading contribution to this is given by
\begin{equation}
\Pi^{(2)}_f(k^2)=\frac{2\al}{\pi}\left[\frac{1}{3}(1-2m_e^2/k^2) \left(\sqrt{1+4m_e^2/k^2}\,\sinh^{-1}(k/(2m_e))-1\right)+\frac{1}{18}\right] = \frac{2\al}{\pi}\Pi^{(2)}(k^2)\,.
\end{equation}
To calculate the $\Pi^{(2)}_f(k^2)/k^2$ correction, it is only necessary to introduce $\Pi^{(2)}_f(k^2)$ into Eq.\,(\ref{del0}). The result, after rescaling, is
\begin{equation}\label{del2}
EVP_\de = \frac{\mu^3\,Z\al^5}{2\pi^2 m_1^2} \int_0^\infty\!\!dy\,y^2\,\,\Pi^{(2)}(y/\be)\rho_{n\ell}(y)\,,
\end{equation}
with $\be=m_ea$. The numerical results are given in Table \ref{FSVP} for $n=2$.

With a slight modifications of the definition of $\rho_{n\ell}$, the other fine structure terms can be treated in a similar way. The expectation value of the fine structure spin-orbit term is
\begin{eqnarray}
\langle n\ell j|V_{LS}|n\ell j\rangle &=& \frac{Z\al}{2}(j(j+1)-\ell(\ell+1)-3/4)\left(\frac{1}{2m_1^2} +\frac{1}{m_1m_2}\right)\int\!\!d^3r\,\frac{|\psi_{n\ell}^m(\vec{r})|^2}{r^3} \nonumber \\
&=& \frac{Z\al}{2(2\pi)^3}(j(j+1)-\ell(\ell+1)-3/4)\left(\frac{1}{2m_1^2} +\frac{1}{m_1m_2}\right)\int\!\!d^3k\,\frac{4\pi}{k^2}\,\tilde{\rho}_{n\ell}(k^2a^2)\,, \label{KNLS}
\end{eqnarray}
where
\begin{equation}
\tilde{\rho}_{n\ell}(k^2a^2)=\int\!\!d^3r \frac{\langle|\psi_{n\ell}^m(\vec{r})|^2\rangle e^{-i\vec{k}\cdot\vec{r}}}{r^2} = \int_0^\infty\!\!dr\,R_{n\ell}^2(r)\frac{\sin(kr)}{kr}\,,
\end{equation}
and the angular brackets denote an average over the degenerate $m$ values. For $n=2$ and $\ell=1$, $\tilde{\rho}_{21}(k^2a^2)$ is
\begin{equation}
\tilde{\rho}_{21}(k^2a^2)=\frac{Z^6}{12a^2(k^2a^2+Z^2)^2}\,.
\end{equation}
Then, after setting $k=y/a$,
\begin{eqnarray}
\langle 2p_j|V_{LS}|2p_j\rangle &=& \frac{\mu^3Z^4\al^4}{12\pi} \left(\frac{1}{2m_1^2} +\frac{1}{m_1m_2}\right)(j(j+1)-11/4) \int_0^\infty\!\!dy\frac{1}{(y^2+1)^2} \nonumber \\
&=& \frac{\mu^3Z^4\al^4}{48}\left(\frac{1}{2m_1^2} +\frac{1}{m_1m_2}\right)(j(j+1)-11/4)\,.
\end{eqnarray}
This agrees with Eq.\,(\ref{fsls}). Because of the $1/r^3$ behavior, the vacuum polarization correction is obtained by inserting $2\al(\Pi^{(2)}+\bar{\Pi}^{(2)})/\pi$ into Eq.\,(\ref{KNLS}), where

\begin{equation}
\bar{\Pi}^{(2)}(k)=\left[\frac{1}{3}-\frac{2m_e^2}{k^2}+\frac{2m_e^2}{k^2} \frac{\sinh^{-1}(k/(2m_e))}{\sqrt{(k^2/(4m_e^2))(1+k^2/(4m_e^2))}}\right]\,,
\end{equation}
yielding
\begin{equation}
EVP_{LS}=\frac{Z^4\mu^3\al^5}{6\pi^2}\left(\frac{1}{2m_1^2} +\frac{1}{m_1m_2}\right)(j(j+1)-11/4)\int_0^\infty\!\!dy\,\frac{(\Pi^{(2)}(Zy/\be) +\bar{\Pi}^{(2)}(Zy/\be))} {(y^2+1)^2}\,.
\end{equation}
The remaining fine-structure electron vacuum polarization correction can be obtained in an analogous way by defining a modified $\rho_{n\ell}(k^2a^2)$ that involves averages of derivatives of the radial wave function. Explicitly,
\begin{equation}\label{VSI}
-\frac{Z\al}{2m_1m_2}\langle n\ell|p_i\left(\frac{\de_{ij}}{r}+ \hat{x_i}\frac{1}{r}\hat{x_j}\right)p_j|n\ell\rangle = -\frac{Z\al}{2(2\pi)^3m_1m_2} \int\!\!d^3k\,\frac{4\pi}{k^2}\,\hat{\rho}_{n\ell}(k^2a^2)\,,
\end{equation}
where
\begin{equation}
\hat{\rho}_{n\ell}(k^2a^2)=\frac{1}{k}\int_0^\infty\!\!dr\left[2r\left(\frac{dR_{n\ell}(r)}{dr}\right)^2 +\frac{\ell(\ell+1)}{r}R_{n\ell}^2(r)\right]\sin(kr)\,.
\end{equation}
The vacuum polarization correction is obtained by inserting $\Pi_f^{(2)}(k^2)$ into the integral in Eq.\,(\ref{VSI}). Table \ref{FSVP} contains the numerical results. 

In addition, there are second order perturbative corrections that involve $VP(r)$ and one of the terms contributing to the fine-structure corrections. They all can be reduced to the form
\begin{equation}
E^{(2)}_{n\ell}=2\mu^2Z\al\int_0^\infty\!\!dr\,r^2R_{n\ell}(Zr/a)VP(r)\int_0^\infty\!\! dr'\,r'^2\,g_{n\ell}(Zr/a,Zr'/a)V_{FS}(r')R_{n\ell}(Zr'/a)\,.
\end{equation}
The various contributions are enumerated in Table \ref{FSVP}.
\begin{center}
\begin{table}[h]
\begin{tabular}{|l|r|r|r|}
\hline
  &$2s_{1/2}$ &$2p_{1/2}$ &$2p_{3/2}$ \\
\hline
\hh $\de(\vec{r})$ VP      & $0.872$ & $-0.021$ & $-0.021$ \\
\hh Spin-Orbit VP          & $0.000$ & $-0.088$ & $0.044$  \\
\hh Spin Independent VP    & $-0.017$& $-0.003$ & $-0.003$ \\
\hline
\hh Total Fine Structure VP& $0.855$ & $-0.112$ & $0.020$ \\
\hline
\hh 2nd Order $\de(\vec{r})$-VP  & $1.403$  & $0.000$ & $0.000$     \\
\hh 2nd Order $(\vec{p}^{\,2})^2$-VP & $-1.058$ & $-0.065$ & $-0.065$ \\
\hh 2nd Order Spin-Independent-VP& $-0.083$ & $-0.007$ & $-0.007$ \\
\hh 2nd Order Spin-Orbit-VP         & $0.000  $ & $-0.096$ & $0.048$  \\
\hline
\hh Total 2nd Order Fine Structure VP&$ 0.262$ & $-0.168$&$-0.024$\\
\hline
\end{tabular}
\caption{Electron vacuum polarization corrections to the $n=2$ states of muonic helium are shown. The entries are in meV. \label{FSVP}}
\end{table}
\end{center}

\section{Details of the Vacuum Polarization Corrections to the Nuclear Size}

Using the nuclear charge form factor model $F(k^2)=(r_N^2 k^2/12 +1)^{-2}$, the inclusion of the vacuum polarization correction is given by
\begin{equation}
\Delta E_N(n\ell)=-\frac{2\mu Z\al^2}{\pi}\int\!\!dx\,\Pi^{(2)}_f(x/\be) \left(F(x/a)-1\right)\rho_{n\ell}(x^2)\,.
\end{equation}
For $n=2$, $\ell=0,1$ numerical integration gives the results in Table \ref{NVP}.

The second order vacuum polarization correction to the nuclear size for $n=2,\ell=0$ can be written
\begin{equation}
\Delta E_N=2\mu^2Z\al\int\!\!dr\,r^2R_{20}(Zr/a)VP(r)\int\!\!dr'\,r'^2g_{20}(Zr/a,Zr'/a)\Delta V_N(r)R_{20}(Zr'/a)\,,
\end{equation}
where $\Delta V_N(r)$, the correction to $-Z\al/r$, is
\begin{equation}
\Delta V_N(r)=\frac{Z\al}{r}\left(1+\sqrt{3}r/r_N\right)e^{-2\sqrt{3}r/r_N}\,.
\end{equation}
The integrals over $r$ and $r'$ can be integrated analytically and the integral over $\De^{(2)}(x)$ was evaluated numerically. The result is found in Table \ref{NVP}. The contribution from the $2p$ state is negligible. 

\begin{center}
\begin{table}[h!]
\begin{tabular}{|l|r|r|r|}
\hline
  &$2s_{1/2}$ &$2p_{1/2}$ &$2p_{3/2}$ \\
\hline
\hh Nuclear Size $\langle r^2\rangle$       & $295.149$ & $0.000$ & $0.000$ \\
\hh Nuclear Size $\langle r^2\rangle^{3/2}$ & $ -5.419$ & $0.000$ & $0.000$ \\ \hline
\hh Total Nuclear Size                      & $292.198$ & $0.000$ & $0.000$ \\
\hline
\hh Nuclear Size $F(q^2)$ VP                & $0.887$   & $-0.023$& $-0.023$
\\ 
\hh 2nd Order Nuclear Size-VP               & $1.429$   & $0.000$ & $0.000$ \\
\hline
\hh Total Nuclear Size-VP                   & $2.316$   & $-0.023$ &$-0.023$\\ \hline
\end{tabular}
\caption{Nuclear Size corrections to the $n=2$ states of muonic helium. The entries are in meV. \label{NVP}}
\end{table}
\end{center}

\acknowledgments
The author would like to thank Professor Randolf Pohl for drawing his attention to Ref.\cite{Pohl} and for pointing out Ref.\cite{Sick} which reports the latest value of the $^4$He nuclear radius.

\end{document}